\input harvmac
$\,$
\overfullrule=0pt
\input epsf
\def\Box{\mbox{.08}{.08}\,}     
\vskip-2cm
\Title{\vbox{\baselineskip12pt\hbox{\phantom{lll}}\hbox{}}}
{\vbox{\centerline{ Global Theory of Quantum Boundary Conditions}\medskip
\centerline {and Topology Change}}}
\vskip-1cm
\centerline{\sevenrm \bf M. Asorey$^a$,  A. Ibort$^b$ and G. Marmo$^c$}
\vskip .1cm

\centerline  {$^a$\sevenrm \it Depto. de F\'{\i}sica Te\'orica,
Universidad de Zaragoza.  50009 Zaragoza, Spain.}
\vskip 0.05cm
\centerline  {$^b$\sevenrm \it Depto. de Matem\'aticas, Univ.
Carlos III de Madrid.  28911 Legan\'es, Madrid,  Spain. }
\vskip 0.05cm
\centerline {$^c$ \sevenrm \it Dipto. di Scienze Fisiche, INFN Sezione di Napoli, Univ. di Napoli Federico II,}
\centerline{\sevenrm \it 
80126 Napoli, Italy.}

\def\bc{boundary conditions}
\def\Tr{\rm Tr}
\def\back{{{\raise.4em\hbox{$\, _\backslash\,$}}}}

 2

\font\tenblack=msbm10
\def\sp #1{{{\cal #1}}}

\def\smallfield #1{\hbox{{\tenblack #1}}}

\def\sp #1{{{\cal #1}}}

\def\set#1{\{\,#1\,\}}
\def\frac#1#2{{#1\over #2}}

\def\big R{{\hbox{{\bigfield R}}}}
\def\bbig R{{\hbox{{\bbigfield R}}}}
\def\H{{\hbox{{\smallfield H}}}}

\font\af=msbm10

\font\bbm=bbm10
\def\Z{\hbox{\af Z}}
\def\R{\hbox{\af R}}

\def\C{\hbox{\af C}}

\def\I{\hbox{\bbm 1}}
\def\id{\hbox{\bbm 1}}

\mathchardef\imath="717B
\def\inbar{\,\vrule height1.5ex width.4pt depth0pt}
\def\IB{\relax{\rm I\kern-.18em B}}
\def\IC{\relax\hbox{$\inbar\kern-.3em{\rm C}$}}
\def\ID{\relax{\rm I\kern-.18em D}}
\def\IE{\relax{\rm I\kern-.18em E}}
\def\IF{\relax{\rm I\kern-.18em F}}
\def\IG{\relax\hbox{$\inbar\kern-.3em{\rm G}$}}
\def\IH{\relax{\rm I\kern-.18em H}}
\def\II{\relax{\rm I\kern-.18em I}}
\def\IK{\relax{\rm I\kern-.18em K}}
\def\IL{\relax{\rm I\kern-.18em L}}
\def\IM{\relax{\rm I\kern-.18em M}}
\def\IN{\relax{\rm I\kern-.18em N}}
\def\IO{\relax\hbox{$\inbar\kern-.3em{\rm O}$}}
\def\IP{\relax{\rm I\kern-.18em P}}
\def\IQ{\relax\hbox{$\inbar\kern-.3em{\rm Q}$}}
\def\IR{\relax{\rm I\kern-.18em R}}
\font\cmss=cmss10 \font\cmsss=cmss10 at 10truept
\def\IZ{\relax\ifmmode\mathchoice
{\hbox{\cmss Z\kern-.4em Z}}{\hbox{\cmss Z\kern-.4em Z}}
{\lower.9pt\hbox{\cmsss Z\kern-.36em Z}}
{\lower1.2pt\hbox{\cmsss Z\kern-.36em Z}}\else{\cmss Z\kern-.4em Z}\fi}
\def\IGa{\relax\hbox{${\rm I}\kern-.18em\Gamma$}}
\def\IPi{\relax\hbox{${\rm I}\kern-.18em\Pi$}}
\def\ITh{\relax\hbox{$\inbar\kern-.3em\Theta$}}
\def\IOm{\relax\hbox{$\inbar\kern-3.00pt\Omega$}}

\def\CC{{\cal C}}

\def\CO{{\cal O}}
\def\CM{{\cal M}}

\def\CH{{\cal H}}

\def\CU{{\cal U}}

\def\tg{{\rm \ tg\ }}
\def\ctg{{\rm \ ctg\ }}
\def\det{{\rm \, det\, }}

\baselineskip=16pt plus 2pt minus 1pt
\bigskip
\centerline{\bf Abstract}
\vskip .3cm
\baselineskip=14pt plus 2pt minus 1pt

We analyze the global theory of  boundary conditions for a constrained 
quantum system with  classical configuration space a compact Riemannian 
manifold $M$ with regular boundary $\Gamma=\partial M$.
The space $\CM$ of self-adjoint extensions of the covariant Laplacian
on $M$ is shown to have interesting geometrical and
topological properties which are related to the different topological
closures of $M$. In this sense, the change of topology of $M$ is connected with
the non-trivial structure of $\CM$. The space $\CM$ itself can be
identified with the unitary group $\CU(L^2(\Gamma,\C^N))$ of the
Hilbert space of boundary data $L^2(\Gamma,\C^N)$. This description,
is shown to be equivalent to the  classical von Neumann's description
in terms of
deficiency index subspaces, but it is more efficient and explicit because it
is given only in terms of the boundary data, which are the natural
external inputs of the system.   A particularly interesting family of
boundary conditions, identified as the set of unitary operators 
which are  singular under the Cayley transform,
$\CC_-\cap\, \CC_+$ (the Cayley manifold), turns out to  play a relevant
role in topology change phenomena.
The singularity of the Cayley transform implies that some energy levels,
usually associated with edge states, acquire an infinity energy when
by an adiabatic change the \bc\ reaches
the Cayley submanifold $\CC_-$. In this sense topological transitions require
an infinite amount of quantum energy to occur, although the description
of the topological transition in the space $\CM$ is smooth. This fact
has relevant implications in string theory 
for possible scenarios with joint descriptions of open and closed strings.
In the particular case of elliptic self--adjoint \bc, the space $\CC_-$
can be identified with a Lagrangian submanifold
of the infinite dimensional Grassmannian. The corresponding
Cayley manifold $\CC_-$  is dual of the Maslov class of
$\CM$. The phenomena are illustrated with some simple low dimensional examples.
\tenrm
 \overfullrule=0pt  \hyphenation{systems}
\bigskip\vfill \noindent\baselineskip=16pt plus 2pt minus 1pt

\Date{ }  
\newsec{Introduction}

The analysis of the role of the  boundary of  quantum systems has 
became a recent focus of activity in different branches of
physics which range from the analysis of edge states
in the Hall effect \ref\ed{Halperin, Phys. Rev. {\bf B 25}(1982) 3529
\semi D.J. Thouless, Phys. Rev. Lett. {\bf 71} (1993) 1873
\semi A.H. MacDonald, Phys. Rev. Lett. {\bf 64} (1990)
220 \semi X.G. Wen, Phys. Rev. Lett. {\bf 64} (1990)
2206; Mod. Phys. Lett. {\bf B5}(1991) 39 ; Int. J.
Mod. Phys. {\bf B6}(1992) 1711 \semi
M. Stone and M.P.A. Fisher,  Int. J.
Mod. Phys. {\bf B6}(1994)2539 } to quantum black hole physics
\ref\bh{S. Hawking, Commun. Math. Phys. {\bf 43} (1975) 199}\ref\bhh{J.
Bekenstein, Phys. Rev. {\bf D7} (1973) 2333},
quantum gravity \ref\qg{G. t' Hooft, gr-qc/9310026, arXiv:gr-qc/9606088
}, cosmology \ref\vil{A. Vilenkin, Phys. Rev.
{\bf D33}(1986) 3560 },
strings, D-branes \ref\pol{J. Polchinski, Phys. Rev. Lett. {\bf 75}(1995)
4724} and M-theory (see \ref\wit{E. Witten, Adv. Theor. Math. Phys.
 {\bf 2}(1988) 253} for a review).
In QFT the relevance of boundary conditions is also crucial for phenomena
like spontaneous breaking of symmetries, anomalies
\ref\anm{N.S. Manton, Ann. Phys. (N.Y.) {\bf 159} (1985) 220
\semi J.G. Esteve, Phys. Rev. {\bf D 34} (1986) 674  \semi M. Aguado, M. Asorey and J.G. Esteve,
Commun. Math. Phys. {\bf 218} (2001)
233},
the  Casimir effect \ref\vac{P. Milonni, {\it The Quantum Vacuum: An Introduction to Quantum Electrodynamics}, Academic Press, San Diego (1994) }
or the analysis of the anysotropic structure of the cosmic
background radiation \ref\dod{J.-P. Luminet, A. Riazuelo, R. Lehoucq and
J.P. Uzan, Nature {\bf 425} (2003) 593}.

The conservation of  probability  in quantum mechanics, which is
intrinsically connected with the unitarity principle,
imposes severe constraints on the boundary behaviour of quantum
states in systems evolving in bounded domains. The analytical condition,
which is encoded by selfadjointness  of the hamiltonian operator,
contains all the quantum subtleties associated
to the unitary principle and the dynamical behaviour at the boundary.
In the classical field physics there are not so stringent conditions and
the classification of the different types of boundary conditions is
basically based on phenomelogical considerations rather than in basic
physical principles.
The existence of a boundary generically
enhances the genuine quantum aspects of the system.
Famous  examples of this behaviour are
Young slit experiments and the Aharanov-Bohm effects, which
pointed out the relevance of  \bc\  in the quantum theory.
Another examples of quantum physical phenomena which
are intimately related to boundary conditions are the Casimir effect
\ref\cas{H.B.G. Casimir, Proc. K. ned. Akad. Wet. {\bf 51}(1948) 793},
the role of edge states \ref\sr{V. John, G. Jungman and  S. Vaidya, Nucl. Phys.
{\bf B 455 } (1995) 505}
and the quantization of conductivity \ref\qcd{ D.J. Thouless, M. Kohmoto, M.
P. Nightingale and M. den Nijs, Phys. Rev. Lett. {\bf 49} (1982) 405\semi
J. Avron and R. Seiler, Phys. Rev. Lett. {\bf 54} (1985)259}
in the quantum Hall effect.  The physics of boundary conditions is
becoming very relevant in quantum gravity, string theory and brane theory.
Effects like topology change \ref\rbal{ A.P. Balachandran, G. Bimonte,
G. Marmo and A. Simoni, Nucl. Phys. {\bf B 446} (1995) 299},
quantum holography \qg\ref\hol{
L. Susskind, J. Math. Phys. {\bf 36} (1995)6377; arXiv:gr-qc/9409089}
and AdS/CFT correspondence \ref\mal{J. Maldacena, Adv. Theor. Phys.{\bf 2}
(1998) 231.} show the relevance of boundaries in the description
of fundamental physical phenomena. Moreover, the recent observation of a
suppression of quadropole and octopole components of the cosmic
background radiation might be connected with the boundary conditions
or the space topology of the Universe \dod.
To some extent the role of  boundary phenomena
has been promoted from  academic and phenomelogical simplifications
of more complex physical  systems to a higher status connected with
very basic fundamental principles.

Another kind of interesting applications arise in pure mathematics in the
study of the index theorem for
manifolds with boundary \ref\aps{M. Atiyah, Patodi and Singer,
Bull. London, Math.
Soc. {\bf 5 }(1972) 229; Proc.
Camb. Philos. Soc. {\bf 77 }(1975)43; {\bf 78 }(1976)405; {\bf 79 }(1977)71}.

The dynamics of a system with boundary requires information about
the physical properties of the boundary. The boundary conditions
macroscopically encode the microscopic or fundamental 
structure of the material medium
that the physical boundary is made off.  In fact the  dynamics
is not well defined until the boundary conditions are not completely specified.
In classical mechanics, boundary conditions determines the evolution
of the system after reaching the boundary. The corresponding
boundary conditions are essentially local, except for those which
correspond to the folding of the boundary and lead to non-trivial
topology changes across the boundary.  In classical field theory boundary
conditions are specified by the values of the fields and some of
its derivatives necessary to solve the corresponding boundary
value problem. In quantum field theories the fluctuations
of the bosonic fields, both  in the bulk and  the boundary, can
contribute to the dynamics of the system for open boundary
conditions, although the nature of the boundary might require
more specific \bc. For fermionic fields
boundary conditions are also needed to guarantee the
consistency of the theory. In gauge theories,
quantum gravity or string theories,
however, a more general type of boundary conditions have to be
considered to describe  the sum over different space-time
topologies.

In  this paper we
consider  the global theory of boundary conditions which are
compatible with
the fundamental properties of a elementary quantum system which is confined
on a bounded domain with boundary.
From a physical point of view the boundary conditions are determined
by the nature of the Hamiltonian in the interior of the domain and by the
physical characteristics of the boundary.
 We analyze the minimal requirements that the quantum
theory imposes on boundary conditions in terms of constraints
on the values of boundary data, and find out all possible
solutions compatible with unitarity.
Among all these boundary conditions one finds those
which correspond to topological foldings of the boundary,
sticky conditions which enhance the role of edge states, and all
kind of classical boundary conditions.
 The space of all these boundary conditions $\CM$ exhibit  very
interesting geometrical and topological structures.
It has a group structure and can be identified with an
infinite dimensional Grassmannian manifold.
The global properties of $\CM$  might be relevant for
quantum gravity where one has to sum over a very large class
of boundary conditions.
The study of such a global properties of $\CM$  and its connection with
the appearance of edge states and topology change is the main
motivation of this paper.  We identify all conditions involving
topology change as a Cayley submanifold of the space of all boundary
conditions. We point out the existence of a connection between topology
change and the existence of edge states with very large negative energies.

We also analyze the connection of this submanifold
with the non-trivial topology of $\CM$ and the Maslow index of the
grasssmanian structure.

In Sect. 2 we derive a  description of quantum boundary
conditions in terms of constraints
on the  boundary data. The equivalence with the
classical von Neumann's description in shown in Sect. 3.
In Sect. 4 we analyze the geometrical structure of the space
of quantum boundary conditions and in Sect. 5, its relation with
the infinite dimensional Grassmannian and the two Cayley submanifolds.
Finally, the appearance of edge states for boundary conditions
in the vicinity of one of the two Cayley submanifolds and its connection with
topology changes is discussed in Sect. 6.

\newsec{Quantum Boundary Conditions}

Let us  consider a non-relativistic point-like particle
moving on  an $n$-dimensional orientable compact Riemannian manifold
$(M,g)$ with smooth boundary\foot{
The theory can be generalized for piecewise smooth boundaries without 
pathological cone singularities.} $\Gamma=\partial M$, under the action of a
a smooth potential $V$ and a gauge field $A$ defined in a hermitian vector 
bundle $E$ over $M$ of rank $N$.

The space of physical states is defined by the completion of
the space of smooth sections $C^\infty(M,E)$ of $E$
with respect to the norm $\parallel . \parallel_2$
induced by the hermitian product
\eqn\riemp { \langle \psi_1, \psi_2 \rangle =
\int_M \, (\psi_1(x), \psi_2(x))_x \, d\mu_g (x),}
where $d\mu_g (x)$ denotes the Riemannian volume form  defined
by the orientation of $M$ and the metric $g$, and $(. , .)_x$
is the hermitian product of $E$ at $x$. In local coordinates,
$d\mu_g (x)=\sqrt{g}\, d^n x$.
The space $L^2 (E)$ of physical states contains also non-smooth
sections and in fact is independent of the vector bundle $E$
considered and can therefore be identified with $L^2(M, \IC^N)$
\ref\as{M. Asorey, Lett. Math. Phys. {\bf 6} (1982) 429}.

Solving the operator ordering problem in an appropriate way yields
to a  quantum Hamiltonian  formally given by
\vskip-.3cm
\eqn\qham{ \H  = - \Delta_A + V,}
\vskip.3cm\noindent
where $\Delta_A =d_{A}^\dagger d_{A}$ is the covariant Laplace-Beltrami
operator and $V$ is the smooth potential on $M$ \foot{The role of the potential $V$
in the discussion is subsidiary provided it  has not singularities.}.
The covariant differential operator $d_{A} \colon C^\infty(M,E) \to \Omega^1(M,E)$ maps
smooth sections of $E$ into 1-forms on $M$ with values on $E$.  
There is a natural inner product on the space of $E$-valued 1-forms
 $\Omega^1(M,E)$ defined as
$$\langle \alpha, \beta \rangle =
\int_M \, (\alpha (x), \beta (x))_x \, d\mu_g (x),$$
with $(\alpha (x), \beta (x))_x = g^{ij}(x) h^{ab}(x) \alpha_i^a (x) \beta_j^b$ where
$h = h^{ab} \sigma_a \otimes \sigma_b$ denotes the metric bundle and $g = g_{ij} dx^i \otimes  dx^j$
the riemannian metric.   We shall denote by
$d_{A}^\dagger$ the adjoint operator of $d_A$ with respect to the Hilbert space 
structure defined by hermitian product above.

It is obvious that $\H$ is a symmetric operator on
$C^\infty_0 (M,E)$, the space of smooth
sections with compact support in the interior of $M$.
However, in general, $\H$ is not even
essentially self-adjoint in $L^2 (E)$ because the domain $C^\infty_0
(M,E)$ is too small, although  dense in $L^2 (E)$.
The adjoint operator  $\H^\dagger$ is given by the extension of
$\H$ to the dense subspace $H^2(E)$ of class 2 Sobolev sections of $L^2 (E)$
defined by the closure of the space of smooth sections $C^\infty (M,E)$ of 
$E$ with
respect to the norm associated to the hermitian product
\eqn\nprod{\langle \psi_1, \psi_2 \rangle_2 =
\int_M \, ( \psi_1(x), (-\Delta_A +I) \psi_2(x))_x \, d\mu_g (x) .}

The obstruction to the self-adjointness of $\H$ is
due to the non-trivial structure of the boundary terms, the Lagrange form,
appearing in the integration by parts formula
\eqn\simet{ \langle \psi_1, \H \psi_2\rangle =
\langle \H \psi_1, \psi_2\rangle + i  \Sigma\left(\psi_1,
\psi_2\right),
}
valid for any pair of
smooth sections $\psi_1,\psi_2\in C^\infty(M,E)$.
The boundary term
\eqn\sig{ \Sigma\left(\psi_1, \psi_2\right) ={i}
\int_M d \left[
\left( \ast d_{A}\psi_1 , \psi_2 \right) -
\left( \psi_1 ,\ast d_{A} \psi_2 \right)\right],}
that by Stokes theorem, takes the form
\eqn \ssig{\Sigma\left(\psi_1, \psi_2\right)={i}\int_{\Gamma} \,
{j}^\ast\left[\left( \ast d_{A}\psi_1 , \psi_2 \right) -
\left( \psi_1 ,\ast d_{A} \psi_2 \right) \right]
= {i} \int_{\Gamma} \,
\left[(\dot\varphi_1 , \varphi_2 ) -
(\varphi_1 ,\dot\varphi_2 )\right] d\mu_\Gamma
}
which only really depends on the boundary values
$$\varphi_i=j^\ast \psi_i=\psi_i|_\Gamma \qquad (i=1,2) ,$$
of $\psi_1$ and $\psi_2$
and its oriented covariant normal  derivatives
$\dot\varphi_1, \dot\varphi_2$
given by
$$j^\ast [\ast d_{A}\psi_i] =\dot\varphi_i \,d\mu_\Gamma \qquad (i=1,2).$$
We denote by
$\ast$  the Hodge star operator of the orientable
Riemannian manifold $(M,g)$, ${j}^\ast$ is the pullback along the
immersion ${j}:\Gamma\to M$ and $d\mu_\Gamma= i_\nu d\mu_g $
the volume form induced on the boundary by the bulk volume form $ d\mu_g$
and the outward normal $\nu$.

The boundary term $\Sigma(\psi_1,\psi_2)$ has a relevant physical
interpretation. It measures  the net flux of probability  across the
boundary. If the operator $\H $ has to be self-adjoint this flux must
be null: the incoming flux has to be equal to  the outgoing flux,
because the evolution operator $\exp it\H$ in such a case is unitary
and preserves probability.  In fact as it was mentioned before, for
sections of compact support, $\psi_1,\psi_2\in C^\infty_0 (M,E)$,
 it vanishes, implying that the Laplace-Beltrami
operator $\Delta_A$ is symmetric in that domain.  The different
self-adjoint extensions will be defined in linear dense subspaces of
$L^2(E)$ containing  $C^\infty_0 (M,E)$ such that
the boundary term $\Sigma$ vanishes.

The  classification of the different possible self-adjoint extensions
will easily be derived from the cancellation conditions of the
boundary term $\Sigma$.
\medskip
\noindent{\bf Theorem 1:}
{\it The set $\CM$ of self-adjoint extensions of
$\H$ is in one-to-one correspondence with the group of unitary
operators of $L^2(\Gamma,\IC^N)$.}
\medskip
{\it Proof:}
The  boundary term $\Sigma$ has to vanish in the
linear domain of any selfadjoint extension of $\H$. Thus, any
selfadjoint extension is  uniquely characterized
by a maximal isotropic subspaces of  the boundary data  space
$$L^2(\Gamma,\IC^N)\times L^2(\Gamma,\IC^N)=\left\{(\varphi,
\dot\varphi); \varphi,
\dot\varphi\ \in L^2(\Gamma,\IC^N)\right\}$$ 
obtained from the completion\foot{This 
space does not really depends on which
vector bundle $E$  we use to define the hermitian product. 
It only depends on its rank $N$, i.e. $L^2(E)=L^2(\Gamma,\IC^N)$} of
$C^\infty(\Gamma,j^\ast E)\times C^\infty(\Gamma,j^\ast E)
= \left\{(\varphi, \dot\varphi); \varphi, \dot\varphi\ \in C^\infty(\Gamma,
j^\ast E)\right\}$
with repect to the  pseudo-hermitian product (Lagrange form) defined by \ssig\
\eqn\sk{\Sigma\left((\varphi_1,\dot\varphi_1),(\varphi_2,
\dot\varphi_2)\right)=  {i} \int_{\Gamma} \,
\left[(\dot\varphi_1 , \varphi_2 )_x -
(\varphi_1 ,\dot\varphi_2 )_x\right] d\mu_\Gamma.
}
In the space of boundary data
there is also an additional hermitian structure given by
\eqn\bph{\left<(\varphi_1,\dot\varphi_1),(\varphi_2,
\dot\varphi_2)\right>=\langle\varphi_1,\varphi_2\rangle +
\langle \dot\varphi_1,\dot\varphi_2\rangle,}
where
\eqn\bp{\langle\varphi_1,\varphi_2\rangle=\int_\Gamma(\varphi_1 ,\varphi_2 )_x
d\mu_\Gamma}
denotes the hermitian product on $L^2(\Gamma,\C^N)$ defined by the
induced Riemannian structure of the boundary.
The identification  of the  maximal isotropic spaces of the space of boundary
data
becomes easier if we perform the Cayley transform defined by
\eqn\cay{\eqalign{C\left((\varphi_1,\dot\varphi_1);(\varphi_2,
\dot\varphi_2)\right)&=\left((\phi_1^+,\phi_1^-);( \phi_2^+,\phi_2^-)\right)\cr
&=\left((\varphi_1+i\dot\varphi_1,
\varphi_1-i\dot\varphi_1);(\varphi_2 + i\dot\varphi_2,\varphi_2-
i\dot\varphi_2)\right).}}
This transformation is an isometry with respect to the hermitian
structure $\langle .\,, .\,\rangle$ but it does not preserves the pseudo-hermitian
structure $\Sigma(.\,,.\,)$ defined by the boundary terms.

In fact, the Cayley transform maps $\Sigma(.\,,.\,)$ into the
pseudo-hermitian product $\Sigma_c(.\,,.\,)$ given by
\eqn\caycc{ \Sigma_c \left((\phi_1^+,\phi_1^-),(\phi_2^+,
\phi_2^-)\right)=\half\langle\phi_1^+,\phi_2^+\rangle-
\half\langle\phi_1^-,\phi_2^-\rangle .}
Thus, it is obvious that the maximal isotropic subspaces of $\Sigma$ in
 $L^2(\Gamma,\IC^N)\times
L^2(\Gamma,\IC^N)$ are in one--to--one correspondence with the space of unitary
operators $\CU(L^2(\Gamma,\IC^N))$ of the boundary
space $L^2(\Gamma,\IC^N)$, i.e. any vector of a maximal isotropic
subspace is of the form \eqn\Uiso{  \varphi - i\dot\varphi = U(\varphi +
i\dot\varphi),}
where $U$ is an unitary operator of $L^2(\Gamma,\IC^N)$ which is
uniquely associated to such a subspace.
\hfill$\Box$

  With the above characterization, the set
of self-adjoint extensions of $\H $ inherits  the group structure of the
group of unitary operators  $\sp U \left(L^2(\Gamma,\IC^N )\right)$.
This characterization can be extended for a larger class of differential
operators on $E(M,\IC^N)$ like Dirac operators $\dsl_{A}$ \ref\dirac{
G. Grubb, Commun. Math. Phys. {\bf 240}(2003) 240}
and higher  order differential operators like $\Delta^k_{A}$ (see
appendix C) or $\dsl_{A}^k$. The only change is that
the hermitian structure of the boundary data changes as we change of operator
(see examples in appendix C for illustration).

There are two natural boundary conditions where the Cayley transform
becomes singular: Dirichlet ($U = - \I$) and Neumann ($U = \I$)
boundary conditions. But the group of boundary conditions is much larger.
In general, for spaces $M$ of dimension higher than one
$\sp U \left(L^2(\Gamma,\IC^N\right)$ is an infinite dimensional group.
There are two particular subsets of $\sp U \left(L^2(\Gamma,\IC^N\right)$
which give rise to boundary conditions which are very easily expressed in
terms of boundary values. They are defined by
\eqn\cbc{\varphi=A_- \dot \varphi ; \qquad \dot\varphi=A_+  \varphi}
in terms of  hermitian operators $A_\pm$.
The fact  that they define selfadjoint extensions of $\H$ follows from
the unitarity of the following operators
\eqn\insa{U_\pm= \pm { \I - iA_\pm\over \I+ iA_\pm}.}


\newsec{Equivalence with Von Neumann theory of self-adjoint extensions}

The theory of selfadjoint extensions of  symmetric operators
 densely defined  in Hilbert spaces
was developed by  von Neumann \ref\rvn{J. von Neumann. Math. Ann., {\bf 102}
(1929) 49/131.}(see also \ref\calk{J.W. Calkin, Trans. Amer. Math. Soc. {\bf
45} (1939)369}
\ref\krein{M.G. Krein, Mat. Sbornik {\bf 62} 
(1947) 431}).
We shall see that for the operator $\H $ it
leads  to the same results as the approach developed in the previous
sections.

The von Neumann theory is based on two  deficiency
spaces $\sp N_\pm$, which are spanned by the zero modes of the operators
$\Delta_A^\dagger \mp i I$

\eqn\deficiency { \sp N_\pm  = \ker (\Delta_A^\dagger \mp i\I)
 .}
The adjoint operator $\Delta_A^\dagger$ of the covariant laplacian $ \Delta_A$
is defined on the subspace
$$\sp D = \overline{\sp D_0} + \sp N_+ + \sp N_- ,$$
of
$ H^2 (E)$ of $L^2$, which in general is larger that the domain
$\sp D_0 = C^\infty_0(M,E) $ of
definition of $ \Delta_A$. The  von Neumann theorem \rvn\ establishes that
\medskip
\noindent{\bf  Theorem 2 (von Neumann):} {\it There exists a one-to-one
correspondence between
self-adjoint extensions of $\Delta_A$ and unitary operators $U$ from
$\sp N_+$ to $\sp N_-$.}
\medskip
{\it Proof:}
The domain of the self--adjoint extension $\Delta^U$ corresponding to the
operator
$U$ is $\sp D_U=\overline{\sp D_0} + (I+U)\sp N_+$ 
where the operator $\Delta^U$
is defined by
$$ \Delta_A^U \psi = \Delta_A \psi_0 - i \xi_+ +i U\xi_+,$$
for any function of the form
$\psi = \psi_0 + (I+U) \xi_+$, with $\xi_+\in \sp N_+$ and $\psi_0\in \overline{\sp D_0}$.
The essential idea for the proof is the existence of a one-to-one
correspondence
between the extensions of the symmetric operator $\Delta_A$ and the
extensions of its Cayley transform operator
$$ \tilde{U} = \frac{\Delta_A -i\,{\rm \I}}{\Delta_A + i\,{\rm \I}},$$
defined from
$ {\rm ran} (\Delta_A + i\, {\rm\I})$ into
$ {\rm ran}(\Delta_A -i\, {\rm \I})$.
When, the deficiency subspaces do have the same dimension, it is
possible to extend $\tilde U$ to an unitary operator defined in the whole
$L^2(E)$ Hilbert space. The possible different extensions ${\tilde U}_U$
 are parametrized by
the unitary operators $U$ from
${\rm ran}
( \Delta_A + i {\rm \I})^\perp = \ker (\Delta_A^\dagger -i{\rm\I})$ into
${\rm ran} ( \Delta_A  -i {\rm \I})^\perp = \ker
(\Delta_A^\dagger + {i \rm\I})$. To every such extension we can associate
a selfadjoint extension of $\Delta_A$ by the inverse Cayley transformation
\eqn\invcay{\Delta_A=i {{\tilde U}_U-\id\over {\tilde U}_U+\I} ,}
which is now well defined on $\sp D_U$. \hfill$\Box$

\medskip

Let us now analyze the direct connection
of this theory with the approach developed in the previous
sections.  This can be achieved in two steps.

First, we remark that unitary operators from $\sp N_+$ into
$\sp N_{-}$ are in
one--to--one correspondence with maximal isotropic subspaces of
  the total deficiency space $\sp H_\pm
  =\sp N_+
\oplus \sp N_{-}$, with respect to the natural
pseudo-hermitian structure
of $\sp H_{\pm}$ defined by
\eqn\sss{
\Sigma_\pm\left((\psi_1^+, \psi_1^-),(\psi_2^+,\psi_2^-)\right)
= 2 \langle \psi_1^+ ,\psi_2^+ \rangle - 2 \langle \psi_1^-,
\psi_2^- \rangle,}
for any pair of vectors
 $(\psi_1^+, \psi_1^-) ; (\psi_2^+,\psi_2^-) \in {\sp H_\pm}$.

The connection between both approaches is established by
the map $j^\ast_\pm:\sp H_{\pm}\to L^2(\Gamma,\C^N)\times L^2(\Gamma,\C^N)$
which applies $(\psi_+,\psi_-)$ into their boundary values $(\phi_+,
\phi_-)\equiv \left(j^\ast\psi_+ + i\dot
\varphi_+, j^\ast\psi_- - i\dot
\varphi_-\right)$ in $L^2(\Gamma,\C^N)\times L^2(\Gamma,\C^N)$, with
$j^\ast\psi_\pm=\varphi_\pm$
and  $j^\ast[\ast d_A \psi_\pm]=\dot\varphi_\pm \, d\mu_\Gamma$.
The map $j^\ast_\pm$
defines a one--to--one correspondence between $\sp H_{\pm}$
and $L^2(\Gamma,\C^N)\times L^2(\Gamma,\C^N)$ as a consequence
of the following lemma.

{\bf Lemma:} The map $j^\ast_\pm $
establishes an isomorphism between the deficiency space
$\sp H_{\pm}$ and boundary data $ L^2(\Gamma,\C^N)\times L^2(\Gamma,
\C^N)$.

{\it Proof:}
The space $\sp H_{\pm}$ can be identified with the kernel of
the operator $\Delta_A^{\dagger 2} +\I$ because of the identity
$\Delta_A^{\dagger 2}+{\rm \I}=
(\Delta_A^{\dagger } -i \I)(\Delta^{\dagger}_A
+i \I)$.  The map $j^\ast_\pm $ is obviously injective because if two
 sections $\psi_1$, $\psi_2$  of $\sp H_{\pm}$ have the
same boundary values their difference  $\psi_1- \psi_2$ will have
vanishing boundary values, and the only section on the kernel
of $\Delta_A^{\dagger 2} + \I$ with vanishing boundary values 
is the null section $\psi=0$. To prove that $j^\ast_\pm $  is surjective
one has to show that for any boundary value in $(\varphi, \dot \varphi)\in
L^2(\Gamma,\C^N)\times L^2(\Gamma,\C^N)$  there
is a section  $\psi$ on $\sp H_{\pm}$
such that  $j^\ast \psi =\varphi$ and $j^\ast[\ast d_A \psi] = \dot\varphi
\, d\mu_\Gamma$. But this follows from the solution of the boundary value
problem for the differential operator $\Delta_A^{\dagger 2}+\I$. There is
a unique solution of the equation $(\Delta_A^{\dagger 2}+\I)\psi=0$ with
boundary values $j^\ast \psi =\varphi$ and $j^\ast[\ast d_A\psi] = \dot
\varphi\, d\mu_\Gamma$. The uniqueness of the solution follows from the strict
positivity of the selfadjoint operator $\Delta_A^{ 2}+\I$ with
Dirichlet-Dirichlet $\varphi=\Delta_A\varphi=0$ boundary conditions
(see appendix C). The existence of the solution of this generalized boundary
value problem can be derived in a similar way to that of second order
differential operators \ref\ds{N. Dunford, J.T. Schwartz.  {\it Linear
Operators, Part II: Spectral theory, self-adjoint operators in Hilbert
space}.  (1963)}. The solution can be explicitely expressed in terms of
the boundary values $\varphi,\dot\varphi$ by integrals formulas only
involving the kernel $G_0(x,y)$ of the selfadjoint extension of  
$\Delta_A^{ 2}+\I$ with Dirichlet-Dirichlet $\varphi=\Delta_A\varphi=0$ 
boundary conditions,
\eqn\bbvp{\psi(x)=\int_\Gamma\ \varphi(y)\, j^\ast\
[\ast d_A\  \Delta_A G_0(x,y)]
-\int_\Gamma\ d\mu_\Gamma(y)\ \dot\varphi(y)\, [\Delta_A G_0(x,y)],}
where the covariant differential operators $d_A$ and $ \Delta_A$ act on the
argument $y$ of the functions and $G_0(x,y)$ is the solution of the equation
$\left(\Delta_A^{ 2}+\I\right) G_0(x,y)=
\delta(d(x,y))$ with the boundary conditions 
$ G_0(x,y)=0$ and $\Delta_A G_0(x,y)=0$ for any  $x,y\in \Gamma$. 
$d(x,y)$ is the Riemaniann distance defined  on $M$ by
the metric $g$. The formula \bbvp\ provides an explicit
expression for the inverse of the map $j^\pm$ and completes the proof
of the lemma.

\hfill $\Box$

\medskip

The second step is to show that $j^\ast_\pm$ is, in fact,  an isometry
between $(\sp H_{\pm},\Sigma_\pm)$ and boundary and 
$ \left(L^2(\Gamma,\C^N)\times L^2(\Gamma,\C^N),\Sigma_c\right)$..
 Since
$\Delta_A^\dagger \psi_a^+ = i\psi_a^+$, $a=1,2$, we have that
\eqn\isom{\eqalign{
0 =& \langle \psi_1^+, (\Delta_A^\dagger - i)\psi_2^+ \rangle =
\langle \psi_1^+, \Delta_A^\dagger \psi_2^+ \rangle -i \langle \psi_1^+,
\psi_2^+ \rangle \cr
=& \langle \Delta_A^\dagger\psi_1^+, \psi_2^+ \rangle + i \Sigma_c
(\phi_1^+,\phi_2^+) - i \langle \psi_1^+, \psi_2^+ \rangle \cr
=& \langle (\Delta_A^\dagger -i)\psi_1^+, \psi_2^+ \rangle - 2i \langle
\psi_1^+, \psi_2^+ \rangle + i  \Sigma_c
(\phi_1^+ ,\phi_2^+) \cr =& - 2i \langle
\psi_1^+, \psi_2^+ \rangle +i \Sigma_c (\phi_1^+,\phi_2^+),
}}
i.e.
$$2 \langle \psi_1^+, \psi_2^+ \rangle= \Sigma_c (\phi_1^+,\phi_2^+)$$
In the same way we obtain
$$ 2\langle \psi_1^-, \psi_2^- \rangle= \Sigma_c (\phi_1^-,\phi_2^-)$$
Hence,
 the product of two elements
 $(\psi_1^+,\psi_1^-), ( \psi_2^+ ,\psi_2^-) \in \sp H_\pm $,
 equals that of the corresponding elements
  $(\phi_1^+,\phi_1^-), ( \phi_2^+ ,\phi_2^-) \in
    L^2(\Gamma, \IC^N)\times  L^2(\Gamma, \IC^N), $
 i.e.
 $$\Sigma_{\pm} ((\psi_1^+, \psi_1^-),(\psi_2^+,\psi_2^-))=\Sigma_c
 ((\phi_1^+,\phi_1^-), ( \phi_2^+ ,\phi_2^-)),$$
which establishes the isometric character of $j^\ast_\pm$.
The results can be summarized in  the following theorem.
\medskip
{\bf Theorem 3:} {\it The boundary value map $j^\ast_\pm $ defines an
isomorphism from the Hilbert space of deficiency vectors
$\left(\sp H_{\pm}, \Sigma_\pm\right)$ to the Hilbert space of
boundary data} $$ \left (L^2(\Gamma,\C^N)\times L^2(\Gamma,\C^N),
\Sigma_c \right).$$
\medskip
In consequence, the maximal isotropic subspaces of $\sp H_\pm$
are mapped into those of  $L^2(\Gamma, \IC^N)\times  L^2(\Gamma, \IC^N)$  
and viceversa.
Moreover, the unitary operators from $\sp N_+$ into $\sp N_-$ are in one
to one correspondence with those of $L^2(\Gamma, \IC^N)$.
This shows the equivalence between the von Neumann theory and
the geometric theory based on boundary data.

\newsec{Selfadjoint extensions, boundary data and
Cayley submanifolds }

The  characterization of selfadjoint extensions of
$\H$ in terms of unitary operators of $\CU(L^2(\Gamma, \IC^N))$, although
equivalent to von Neumann characterization, it
is more useful for physical applications because it is
purely formulated in terms of boundary data. The constraints
involved in the definition of the domain of $\H^U$ imply that the
 boundary values $\varphi$, $\dot\varphi$ of the functions of
such a domain  satisfy the condition
\eqn\Ucond{  \varphi - i\dot\varphi = U(\varphi + i\dot\varphi) .}
Generically, the constraint equation can be solved to express
$\dot\varphi$ as a function of $\varphi$
\eqn\dir{\dot\varphi=-i {\I-U\over \I+U} \varphi}
or, alternatively,
$\varphi$ as a functions of $\dot\varphi$
\eqn\adir{\varphi=i {\I+U\over \I-U}\dot\varphi.}
This explicit resolution of the constraint on the boundary data
means that unitarity requires that only half of the dynamical data
are independent on the boundary.

The equations \dir\adir\ are in fact two different expressions of the
Cayley transform relating selfadjoint  and unitary operators
\eqn\cayley{A=-i {\I-U\over \I+U}\qquad A^{-1}=i {\I+U\over \I-U}.}
The inverse transformation being also a Cayley transform
\eqn\invcayley{U= {\I-iA\over \I +iA}.}

It is obvious that $A$ is only well defined if and only if $-1$
does not belong to the spectrum of $U$.
The existence of  $A^{-1}$ requires that the spectrum of $U$
does not contain the unit $1\notin \sigma(U)$.

The previous considerations show that there is a distinguished set of
self-adjoint extensions of $\H$ for which the expression of the
boundary conditions defining their domain cannot be reduced to the
simple form given by eq. \dir\ or \adir.  These
self-adjoint extensions correspond to the cases where $\pm 1$ are in the
spectrum of the corresponding unitary operator
The Cayley submanifolds $\sp C_\pm $
are the subspaces of self-adjoint extensions
which cannot be defined from $\dir$ or $\adir$ and they
can be described equivalently as follows:
\eqn\cayleysur{ \sp C_\pm = \left\{U\in \sp U\left( L^2 (\Gamma,\C^N)\right)\Big| \pm 1 \in \sigma (U)\right\} .}

Notice that the unitary operators $U=\pm \I$ are in the Cayley submanifolds
$\sp C_\pm$, respectively.  $U=-\I$ belongs to the Cayley submanifold
$\sp C_-$ and corresponds to
Dirichlet boundary conditions \adir:
\eqn\dirich{  \varphi = 0 .}
 $U=\I$ is not in the Cayley submanifold
$\sp C_-$ but in $\sp C_+$  and corresponds to the self-adjoint
operator $A = 0$ which defines Neumann boundary conditions
{\eqn\neuma{  \dot\varphi = 0 .}

There is a formal property which distinguishes the two Cayley submanifolds. The
submanifold $\CC_+$ has a group structure whereas $\CC_-$ and thus also
$\CC_-\cap\,\CC_+$ does not because the composition is not a inner operation.

\newsec{The self-adjoint Grassmannian}

The identification of the space $\CM$
with the unitary group of boundary data $\CU(L^2(\Gamma,\C^N))$
provides a group structure to the space of
 selfadjoint realizations of $\Delta_A$.
This also shows that it has a non-trivial topological structure.
All even homotopy groups  vanish
$\pi_{2n}(\CM)=0$ but all odd homotopy
groups are non-trivial $\pi_{2n+1}(\CM)=\Z$  (Bott periodicity theorem).
The fact that the first homotopy group
$\pi_1(\CM)=\Z$ means that the space of boundary conditions is
non-simply connected.  However the set of selfadjoint operators
in  $L^2(\Gamma, \IC^N)$ is a topologically trivial manifold.  This means that
the characterization of selfadjoint extensions of $\Delta_A$
by means of the Cayley transform \dir\ or \adir\ cannot provide
a global description of $\CM$. In fact, the parametrization
\invcayley\ and its inverse
 \eqn\invcayleyy{U^{-1}= {\I + iA\over \I - iA}.}
can be considered as local coordinates in the  charts $\CM \back \CC_\pm$
of the space $\CM$  of selfadjoint  extensions $\Delta_A$.
The topology of each chart is trivial but that of $\CM$ is
highly non-trivial. In this sense, the Cayley submanifold $\CC_\pm$
intersects all non-contractible cycles of $\CM$.

Since $\pi_0 (\CM ) = 0$ and $\pi_1(\CM) = \Z$ the first cohomology group
of $\CM$ is $H^1(\CM)=\Z$. The generator of this cohomology group
is given by the first Chern class of the determinant bundle defined over 
$\CM$. The determinant of infinite dimensional operators $U$ is ill
defined and proper definition requires the introduction of an 
ultraviolet regularization.
In particular, it is  necessary to restict the boundary conditions
to the subspace $\sp M'$ defined by the  unitary $U$ operators of
$\sp M$ which are of the form $U=\I + K$ with $K$ a  Hilbert-Schmidt
operator (i.e. $ \tr\ K^\dagger K< \infty$).
 If $-1\notin\sigma (U)$ this property of $K$
 is equivalent to the requirement that the boundary
operator A is also a Hilbert-Schmidt operator. Indeed, 
$$K_A = \frac{2A}{i\I - A},\quad A={i K_A\over 2+K_A}$$
hence,
$$K_A^\dagger K_A = \frac{4A^2}{\I + A^2} , \quad A^\dagger A={K_A^\dagger K_A
 \over(2+K_A^\dagger) (2+K_A)}$$
and
$$ \Tr\  K_A^\dagger K_A = 4\, \Tr\  \frac{A^2}
{\I + A^2} \leq 4 \Tr\ A^2, \quad \ \Tr\ A^\dagger A \leq  
\Tr\  K_A^\dagger K_A.$$
With this restriction the determinant of $U_A\in \CM'$ can be defined 
by using the standard renormalization
prescription for  determinants
$$\log \det^\prime U = \sum_{i=1}^\infty d_i\,\log\  \frac{1+k_i}{e^{k_i}} ,$$
in terms of   the eigenvalues of $K_A$, $k_i, i=1,2,\cdots $,
and their degeneracies, $d_i, i=1,2,\cdots$.
Finiteness of this prescription for the
 regularized determinant  $\det^\prime U$ follows from the Hilbert-Schmidt
character of $K_A$ which in particular implies a
 discrete spectrum with finite degeneracies 
satisfying the Hilbert-Schmidt condition 
$K_A^\dagger K_A=\sum_{i=1}^\infty d_i |k_i|^2\leq \infty$.

The first Chern class of the regularized determinant bundle
is given by the one form
\eqn\chc{\alpha= \frac{1}{2\pi}\, d\,
\log\, \det^\prime (\gamma(\theta))}

For any closed curve $\gamma \colon S^1 \to \sp M'$
in the
self-adjoint grassmannian,
we  define its  Maslov index $\nu_M(\gamma )$ as the winding
number of the curve
$\det^\prime \circ \gamma \colon S^1 \to U(1)$ \ref\arn{V. Arnold, {\it
Mathematical Methods of Classical Mechanics, Mir, Moscow (1974)}},
in other words,
\eqn\maslow{\nu_M (\gamma) = \frac{1}{2\pi} \int_0^{2\pi}\partial_\theta
\,\log\, \det^\prime (\gamma
(\theta)) d\theta.}
  The Maslow index   $\nu_M(\gamma )$ is the sum of the winding numbers
of the maps 
$\lambda_i(\theta):S^1\to U(1)$ 
described by the flow of eigenvalues of $\gamma$
around $U(1)$. By continuity of $\gamma$ and compactness of $S^1$ it 
follows that only a finite number of eigenvalues reach the value 
$\lambda_i=-1$ for any value of $\theta\in [0, 2\pi)$. It is clear 
that the winding number of the  map $\lambda_i(\theta)$ is
measured by  $\frac{1}{2\pi} \int_0^{2\pi}\partial_\theta
\log\,  (\lambda_i
(\theta)) d\theta$ and also by
the number of indexed crossings of the point $\lambda_i=-1$.
By construction $\nu_M (\gamma)$ is  the finite sum of the 
non-trivial  winding numbers and is always an integer.
This fact and the  existence of  curves with only one crossing
through $-1$ implies that $\alpha$ is in  the generating class of 
the cohomology group
$H^1(\CM',\IZ)$.

The subspace  $\CM'$ of unitary operators of the form  $U=\I + K$
 has richer topological and geometrical structures. In particular
we will see that it is a Grassmaniann, the selfadjoint Grassmannian.


It is obvious that the subspaces
$\CM_+ = L^2 (\Gamma,\C^N) \times \{ {\bf 0} \}$ $= \set{(\varphi, {\bf
0}) \mid \varphi \in L^2 (\Gamma,\C^N)}$ and $\CM_- = \{ {\bf 0}
\} \times L^2 (\Gamma,\C^N)$ $= \set{({\bf 0}, \dot\varphi) \mid
\dot\varphi \in L^2 (\Gamma,\C^N)}$ which correspond to
Dirichlet and Neumann boundary conditions
are isotropic in $\sp M$  and they are paired by
$\Sigma$.  In fact,
$$\Sigma (\varphi_1,{\bf 0}; {\bf 0},\dot{\varphi}_2) = -{i}
\langle \varphi_1, \dot{\varphi}_2 \rangle_{ \Gamma}  $$
The block structure of $\Sigma$ with respect to the isotropic
polarization $ \CM_+ \oplus \CM_-$ of $\sp \CM$ reads
$$\Sigma = \pmatrix{ 0 & -{i} \langle
.\,,.\,\rangle_{\Gamma} \cr
 {i} \langle
.\,,.\,\rangle_{\Gamma} & 0} .$$
Notice that in higher dimensions  only 
the subspace $\sp M_- \subset \sp M$ belongs to  $\sp M'$,
${\sp M_+ }\cap {\sp M'}=\emptyset$.
 
The pseudo-hemitian  
structure $\Sigma$ can be diagonized by means of the Cayley transform
\eqn\cayley{ C (\varphi, \dot\varphi ) =(\phi^+, \phi^- )=
(\varphi + i \dot\varphi, \varphi - i \dot\varphi).
}
which transforms $\Sigma$ into 
$$\Sigma_c= \pmatrix{ \I & 0  \cr
 0 & -\I} .$$

There is  another canonical hermitian 
product on $ \CM_+ \oplus \CM_-$
given by 
$$ \pmatrix{ \I & 0  \cr
0 & \I} $$
which defines a Hilbert structure $\langle .\, , .\,\rangle$ 
on $ \CM_+ \oplus \CM_-$.

The  Grassmannian  $Gr (\CM_+,\CM_-)$ of $ L^2 (\Gamma, \C^N)
\times L^2 (\Gamma,\C^N)$  is the infinite dimensional Hilbert
manifold of closed subspaces $W$ in $ \CM_+ \oplus \CM_-$ such that
the projection on the first factor $\pi_+ \colon W \to \CM_+$ is
a Fredholm operator and the projection on the second factor $\pi_- \colon  W
\to \CM_-$ is Hilbert--Schmidt, that is, \Tr\ $\pi_-^\dagger\pi_- < \infty$
.

The  selfadjoint Grassmaniann  $Gr (\CM_+,\CM_-)\cap\,\CM$
is defined by the selfadjoint extensions of $\Delta_A$ which belong
to the Grassmaniann $Gr (\CM_+,\CM_-)$.  This subspace might be
considered as the space of mild self-adjoint extensions of $\Delta_A$
whose projection into the subspace $\CM_-$ is Hilbert-Schmidt.
It is possible to see that the self-adjoint Grassmannian is a
submanifold of the Grassmannian itself and can be identified with $\CM'$
the space of unitary operators of $\CM$ which are of the 
form $U=\I+K_A$. This follows from the fact that in some parametrization
of $\CM'$
\eqn\marv{\pi_-=\frac{iK_A}{2\sqrt{U}},}
which implies that 
$\Tr\ \pi_-^\dagger \pi_-=\frac{1}{4}\Tr\  K_A^\dagger K_A$, i.e.
 $\pi_-$ is Hilbert-Schmidt if and only if $  K_A$ is Hilbert-Schmidt.

The intersection of the Cayley submanifold $\CC_\pm$
with $\CM'$ defines a subspace of the self-adjoint
Grassmannian $\CC'_\pm \subset \CM'$  which has a stratified structure
according to the number of eingenvalues $\pm 1$ of the corresponding
unitary operator, i.e.
$$\CC'_\pm=\bigcup_{n=1}^{\infty} \CC^{'n}_\pm,$$
 where
$\CC^{'n}_\pm=\{U\in \CU(L^2(\Gamma,\C^N); \pm 1 \in  \sigma (U)\
{\rm with\  multiplicity}\  n\}$.
Notice that the  spectrum of unitary operators in the selfadjoint
Grassmannian is discrete.

Given a continuous curve $\gamma \colon [0,1] \to \CM'$ we 
define its Cayley index $\nu_c(\gamma)$ as the indexed sum of 
crossings of $\gamma$ through the Cayley submanifold $\CC'_-$ 
(notice that the Cayley
submanifolds $\CC'_\pm$ are
oriented).
This is equivalent to the
 the sum of anti-clockwise  crossing of eigenvalues of
$\gamma$ through the point $-1$ on the unit circle
$U(1)$ minus the sum of clockwise crossings weighted with
the respective degeneracies. Therefore, the Cayley index
$\nu_c(\gamma)$  of $\gamma$  is 
equivalent to its  Maslow index $\nu_M(\gamma)$ and we have the
following theorem.
\medskip

{\bf Theorem 4:} {\it The Maslow  and  Cayley indices of a closed curve $\gamma$ in
the selfadjoint Grassmannian are the same
  $\nu_M(\gamma) = \nu_c(\gamma)$.
The Cayley manifold $\sp C'_-$ is dual of the Maslow
class $\alpha$.}
\medskip

For any  unitary operator $U\in \CM$
we will define its degenerate dimension as
the dimension of the eigenspace with eigenvalue $-1$.  If $U$ is in
the self-adjoint Grassmaniann $\CM'$ the dimension of the eigenspace  
with  eigenvalue $-1$ is finite and the
degenerate dimension of the operator is finite.  We shall denote such
number by $n(U)$. It is an  indicator of the level of $\gamma(\theta)$
in  statratification structure 
of $\CC'$: $U=\gamma(\theta)\in\CC^{'n}$ if and only if $n(U)=n$
 The Cayley index of  any curve $\gamma\in \CM'$ 
can be given in terms of this number by the expression
\eqn\rnu{\nu_c(\gamma ) = \int_0^{2\pi}\partial_\theta
n (\gamma(\theta) ) d\theta .}
Since  \rnu\ is the integral of a pure derivative it vanishes unless there
is a singularity in the integrand. This only occurs at the jumps of
$n(\gamma(\theta))$ i.e.  when one more eigenvalue of 
$U=\gamma(\theta)$ becomes equal to $-1$. $\nu_M(\gamma)$ is in fact
a bookkeeping of the  number of eigenvalues of $\gamma(\theta)$ that cross 
through $-1$ and since it is of bounded variation on  
$\CM'$  the integral in eq. \rnu\ is always finite and gives the Cayley
index. This construction provides an alternative 
(singular) characterization of the first Chern
class of the determinant bundle $\det_{\CM'}\left(\CM',U(1)\right)$
and the generating class of the first homology group $H^1 (\CM', \Z)$ of 
$\CM'$.

\newsec{ Topology change and edge states.}

Although the operator $\Delta_A+\I$ is positive in $C^\infty_0(M,E)$
its selfadjoint extensions might not be definite positive.
In fact, if the selfadjoint extension does not belong to any of the Cayley
submanifolds $\CC_\pm$ it is easy to show by integration by parts that
$$\left(d\ \Psi_1,d\ \Psi_2\right) = \left(\Psi_1,\Delta_A
\Psi_2\right)+\left(\varphi_1,
\dot\varphi_2\right)=
\left(\Psi_1,\Psi_2\right)+  \left(\varphi_1, A
\varphi_2\right)=  \left(\Psi_1,\Delta_A \Psi_2\right)+
\left(A^{-1} \dot\varphi_1,
\dot\varphi_2\right),$$
where
$$\left(d\ \Psi_1,d\ \Psi_2\right)=\int_\Gamma\ d\ \Psi_1 \wedge
\ast d\ \Psi_2 ,$$
and $A$ is given by \cayley.

Thus, only if  $\left(d\ \Psi,d\ \Psi\right)-\left(\varphi, A
\varphi\right)$ for every $\Psi$ is positive the operator
$\Delta_A$ will be positive.
In particular if the boundary operator $A$ is positive it might occur
that the whole operator $\Delta_A$ might loose  positivity.
The existence of negative energy levels is thus possible for some
boundary conditions. It can be seen that the states which have
negative energy are in a certain sense edge states.
\medskip
{\bf  Theorem 5:} {\it For any selfadjoint extension  $\Delta_A^U$  of 
$\Delta_A$ whose
unitary operator $U$ has one eigenvalue $-1$ with smooth eigenfunction,
the family of selfadjoint extensions of the form $U_t=U {\rm e}^{i t}$ 
with $t\in (0,\pi/2)$, has for
small values of $t$ one negative energy level which corresponds to an 
edge state. The energy of this edge state becomes infinite when $t\to 0$.}
\medskip
{\it Proof:}
Let  $\xi \in L^2(\Gamma,\IC^N)$ be a smooth eigenstate
of $U$ with eigenvalue $-1$. Then, $U_t\xi={\rm e }^{i t} \xi$.
Let us consider Gaussian coordinates in a collar $\CC_\Gamma\subset M$
around the boundary $\Gamma$ of M. One of those coordinates is the
``radius'' $r$ and the others can identified with boundary
coordinates sifted inside the collar; i.e. $\CC_\Gamma \approx
[1-\epsilon,1]\times \Gamma$. In this coordinates the metric matrix
looks like
\eqn\gmet{g=\pmatrix{1 & 0\cr 0 &\Omega(r,\Gamma) \cr}.}
We shall  consider the following change of coordinates
$r\leftrightarrow s$ with $s={\pi\over 2 \epsilon}(1-r)$.

If we extend the function $\xi$ from the boundary $\Gamma$ to an edge
state $\Psi$ in the bulk domain $M$ by
\eqn\ext{\Psi(x)=\cases{\xi(\Gamma) {\rm e}^{-k\tg s}
{\quad}& $x=(s,\Gamma)\in \CC $ \cr
0 & $x\notin \CC$\cr},}
it is easy to check that the extended  function $\Psi$ is smooth in $M$
and for
$$k={2\epsilon\over\pi}\ctg {t\over 2}$$
belongs to the domain of the selfadjoint extension of
$\Delta_A^{U_t}$ associated to the unitary matrix $U_t=
{\rm e}^{i t} U$.
Thus, we have
\eqn\bcc{\left(\Psi,\Delta_A^{U_t}
\Psi\right)= \left(d_A\, \Psi,d_A\, \Psi\right)-{\rm ctg}\ {t \over 2}
\left(\xi, \xi\right)}
where
\eqn\dd{\eqalign{\left(d_A\, \Psi,d_A\, \Psi\right)=& \int_{0}^{\pi/2}
 ds\ \int_\Gamma\
d\mu_\Gamma(s)\  |\xi |^2  ({k^2\pi\over 2\epsilon})
(1+(\!\tg s)^2)^2{\rm e}^{-2k\tg s} \cr
+& {2 \epsilon \over \pi}\int_{0}^{\pi/2}  ds \ \int_\Gamma\
d\mu_\Gamma(s)\ (\xi^\ast ,\Delta_\Gamma \xi)\  {\rm e}^{-2k\tg s}.
\cr}}
For small enough $\epsilon<<1$ we have that the dependence
on $s$ of $\Omega$ might be negligible $|\Omega(s,\Gamma)| <
|\Omega(0,\Gamma)|(1 + \delta) $. Thus,
\eqn\dzd{{\left(d_A\, \Psi,d_A\, \Psi\right)< \left({k\over 2} + 
{1\over 4 k}
\right) \,{\pi(1 + \delta)\over  2\epsilon} \, \|\xi\|^2 +
{2 \epsilon (1+ \delta)\over \pi } (\xi^\ast , \Delta_\Gamma
\xi)}}
and
\eqn\bbc{{\left(\Psi,\Delta_A^{U_t}
\Psi\right)< {\pi \over  2\epsilon}\left(  {1\over 4 k } (1+ \delta)-
{k\over 2}\, (1- \delta)\right)\,
\|\xi\|^2 +  {2 \epsilon  (1+ \delta) \over \pi } (\xi^\ast,
\Delta_\Gamma \xi),}}
which shows that $\left(\Psi,\Delta_A^{U_t} \Psi\right)<0$ is not positive for
small values of $\varphi=2\, {\rm arc\,ctg}\,(k \pi/2\epsilon)$. Notice that
the normalization of the edge state $\Psi$
$$||\Psi||^2 = \int_M (\Psi^\dagger,\Psi)_x \, d\mu_g (x) \geq
{\pi(1 - \delta)\over  2\epsilon} \, ||\xi||^2\,\int_0^{2\pi}\, ds\,
{\rm e}^{-2 k \tg s}$$
vanishes in the limit $t \to 0$ but it is always
a positive factor for $t \neq 0$ which preserves the bound \bbc.
Moreover, the nature of the edge state $\Psi$ also shows the existence
of a ground state $\Psi_0$ with negative energy which is an edge state.
The energy $E_0$ of this state goes to  $-\infty$ as $t \to 0$,
whereas the edge state $\Psi_0$ schrinks to the edge disappearing from
the spectrum of $\Delta_A^{U_t}$ in that limit.

\hfill$\Box$

Although the role of boundary conditions in the two Cayley submanifold
$\CC_\pm$ is quite similar from the mathematical point of view
the boundary conditions are quite different from the physical
viewpoint. In particular, an analysis along the lines of the proof
of the above theorem leads to the same \bbc\ inequality but with
$$k={2\epsilon\over\pi}\tg {t \over 2}$$
which points out the existence of  edge states with very large
(positive) energy as $t \to 0)$. It can also be shown that
in that limit one energy level crosses the  zero energy level
becoming a zero mode of the Laplacian operator.
Therefore, the role of boundary conditions in $\CC_-$ (e.g. Dirichlet)
is very different of that of boundary conditions in $\CC_+$ (e.g. Newmann).

Notice that the result of the theorem does not require $U$ to be
in the selfadjoint Grassmannian $\CM'$. This is specially interesting,
because there is a very large family of boundary conditions which
do not belong to $\CM'$. In particular, boundary conditions implying
a topology change in higher dimensions
are not in  $\CM'$ because  the corresponding
unitary operators in $\CU(\Gamma)$ present an infinity of  eigenvalues
$\pm 1$ which implies that $U$ cannot be of the form $\I + K$ with
$K$ Hilbert-Schmidt.
 Indeed, all boundary conditions which involve a change of topology, i.e.
gluing together domains $\CO_1$, $\CO_2$ of the boundary $\Gamma$, 
belong to $\CC_-\cap\, \CC_+$.
This property follows from the fact that the boundary conditions imply
that the boundary values $\varphi,\dot\varphi$ are related in the domains
that are being glued together, i.e. $\varphi(\CO_1)=\varphi(\CO_2),
\dot\varphi(\CO_1)=-\dot\varphi(\CO_2)$, respectively.   These 
requirements imply that the unitary operator $U$ corresponding
to this boundary condition is identically $U=\I$ on the subspace of
functions such that
$\varphi(\CO_1)=\varphi(\CO_2)$ and $U=-\I$ on the subspace of
functions such that $\varphi(\CO_1)=-\varphi(\CO_2)$. 
Since both subspaces
are infinite-dimensinal for manifolds $M$ with more that one dimension
it is clear that those operators $U$ do not belong to $\CC'_-\cap\, \CC'_+$.
However the result of Theorem 5 implies that there always exists a
boundary condition close to one involving the gluing of
the domains with very large negative
energy levels. This means that Cayley manifold $\CC_-\cap\,\CC_+$ 
is also very special and that topology change involves an interchange of an
infinite amount of quantum energy. The result might have relevant
implications in quantum gravity and string theory.

\newsec{Conclusions}

We have not analysed asymptotic boundary conditions which appear
in singular boundary problems like a particle moving in a Dirac
delta potential in the plane \ref\dl{R. Jackiw, in {\it
M.A.B. B\'eg Memorial Volume}, eds. A. Ali and P. Hoodbhoy,
World Sci., Singapore (1992) \semi
C. Manuel and R. Tarrach, Phys. Lett. {\bf B 301} (1993) 72 \semi
J. G. Esteve,Phys. Rev. {\bf D66}(2002) 125013 }
or in the asymptotic Minkowskian boundary of anti-deSitter space-times.
The later is of relevant interest in the analysis of the Maldacena conjecture
\mal. However, if we regularize the boundary we can
use the standard boundary conditions discussed throught the
paper and consider some renormalization group flow
limit which keep the selfadjoint
character of the Laplace-Beltrami operator.
Another interesting boundary effects which are  not analysed in this
paper are the deformation of the boundary  and the inclusion of
local insertions. In two dimensions, both effects
are connected with theory of integrable systems (see Refs.
\ref\rpal{J. Palmer, M. Beatty, C. A. Tracy,  Commun. Math. Phys.
{\bf 165} (1994)97}\ref\itep
{A. Marshakov, P. Wiegmann, A. Zabrodin, Commun. Math. Phys. 227 (2002)
131-153})

For smooth boundaries we have given a very simple characterization of
the self-adjoint extensions of the covariant Laplacian in terms of
boundary data, which although equivalent to von Neumann's characterization
based on  deficiency index spaces, is more convenient for physical
applications. The space of all boundary conditions exhibits a
natural group structure and a non-trivial topology. In this space
processes involving topology change can be naturally described.
Boundary conditions involving topology change can describe in the
same scenario open and closed strings and smooth
interpolations between both type of strings. This might be relevant 
for new developments in string theory.
We have shown that for any adiabatic change of boundary conditions which
involves a  crossing of the Cayley submanifold $\CC_-$ there is  an
edge state which becomes an infinite negative energy level
at the boundary condition
on $\CC_-$.
Negative energy states are only possible if they are localized near the
the boundary (edge states).
The number of crossings of $\CC_-$ for any closed loop $\gamma$ of boundary
conditions defines a  Maslov index $c_M(\gamma)$ which identifies
with the generator of the first cohomology group of the space of all boundary conditions $H^1(\CM',\Z)$.

 In the case of the Cayley submanifold $\CC_+$ a
similar argument shows that for any one-parameter family of boundary
conditions which intersects at
the Cayley submanifold $\CC_+$ there is one energy level of the $\Delta_A$
which becomes a zero level state on the Cayley submanifold.

From the physical point of view the above results show that the boundary
has a contribution to the energy levels and some of those energy levels
are rather localized at the edge of the boundary (edge states).
For boundary conditions which  approach a topology changing 
boundary condition at least one energy level acquires a very large negative 
energy which means that such
a transition from the energetic point of view is singular. However, looking
at any time to the corresponding boundary condition no singularity is shown
in the spectrum because those states which become highly energetic also 
become
very singular and disappear from the domain of the selfadjoint operator
and consequently their energy levels
from the spectrum. However, the effect leads to some
observable phenomena. For instance, the dependence of Casimir energy
on the boundary conditions for a scalar field becomes singular  when
the boundary condition approaches a boundary condition with unitary
operators with extra -1 eigenvalues. In particular, this occurs for
boundary conditions involving a topology change which might have relevant
implications for quantum gravity and string theory.

\centerline{ \bf Acknowledgements}
We thank to A.P. Balachandran for discussions. This work has been partially
supported by a grant INFN-CICYT for  cooperation.
The work of M.A. has also been partially supported by the Spanish MCyT grant
FPA2000-1252 as well as A.I. that has been partially supported by the
Spanish MCyT grantS BFM2001-2272 and BFM2002-02888.

\appendix{A}{Topology Change and One Dimensional Boundary Conditions}

To illustrate the utility of the above geometric approach we
consider some simple applications to Sturm-Liouville problems. In
such a case the configuration space is constrained to an interval
$M=[0,1]$ of real numbers.  The metric $g$ is the  standard
Euclidean metric of $\IR$ and the symmetric operator is  the
Sturm-Liouville second order differential operator
$$\IH=\ha\Delta= -\ha{d\over
dx^2}$$ defined on $C^\infty_0([0,1])$. The boundary set is in this
case discrete $\partial\Omega=\{0,1\}$ and
$L^2(\partial\Omega)=\IC^2$. Therefore the different selfadjoint
extension are parametrized by a $2\times 2$  unitary matrix
\eqn\mat{U=\pmatrix{u_{11}&u_{12}\cr
u_{21}&u_{22}}}
The domain of the associated extension is given by the functions of
$ L^2([0,1])$ whose boundary values satisfy the following equations,
\eqn\sturm{\pmatrix{\varphi(0)+i\dot\varphi(0)\cr
\varphi(1)+i\dot\varphi(1)}=\pmatrix{u_{11}&u_{12}\cr
u_{21}&u_{22}}\pmatrix{\varphi(0)-i\dot\varphi(0)\cr
\varphi(1)-i\dot\varphi(1)},}
where $\dot\varphi(0)=\varphi'(0)$ y $\dot\varphi(1)=-\varphi'(1)$.

Some specially interesting examples correspond to the case when the
matrix $U$ is diagonal or {\it anti-diagonal}. In the first case we have
\eqn\diagona{U=\pmatrix{e^{i\epsilon}&0\cr
0&e^{i\gamma}}}
which corresponds to the boundary conditions
\eqn\diric{\eqalign{-\sin
{\epsilon\over 2}\varphi(0)+\cos{\epsilon\over 2}\dot\varphi(0)=0\cr -\sin
{\gamma\over 2}\varphi(1)+\cos{\gamma\over 2}\dot\varphi(1)=0} }
which includes Newmann $\dot\varphi(0)=\dot\varphi(1)=0$ and
Dirichlet $\varphi(0)=\varphi(1)=0$ boundary conditions.
        In the {\it anti-diagonal} case
\eqn\antdia{U=\pmatrix{0&e^{-i\epsilon}\cr
e^{i\epsilon}&0}}
we have (pseudo)periodic boundary conditions
\eqn\per{\eqalign{&\varphi(1)=e^{i\epsilon}\varphi(0)\cr
&\varphi'(1)=e^{i\epsilon}\varphi'(0)} }
$\varphi(1)=e^{i\epsilon}\varphi(0)$
with probability flux propagating through the
boundary. This condition is in fact a topological boundary
condition which corresponds to the folding the interval
into a circle $S^1$, with a magnetic flux $\epsilon$
crossing throughout.
\bigskip
\hskip 2.0cm \epsfxsize=8cm\epsfbox{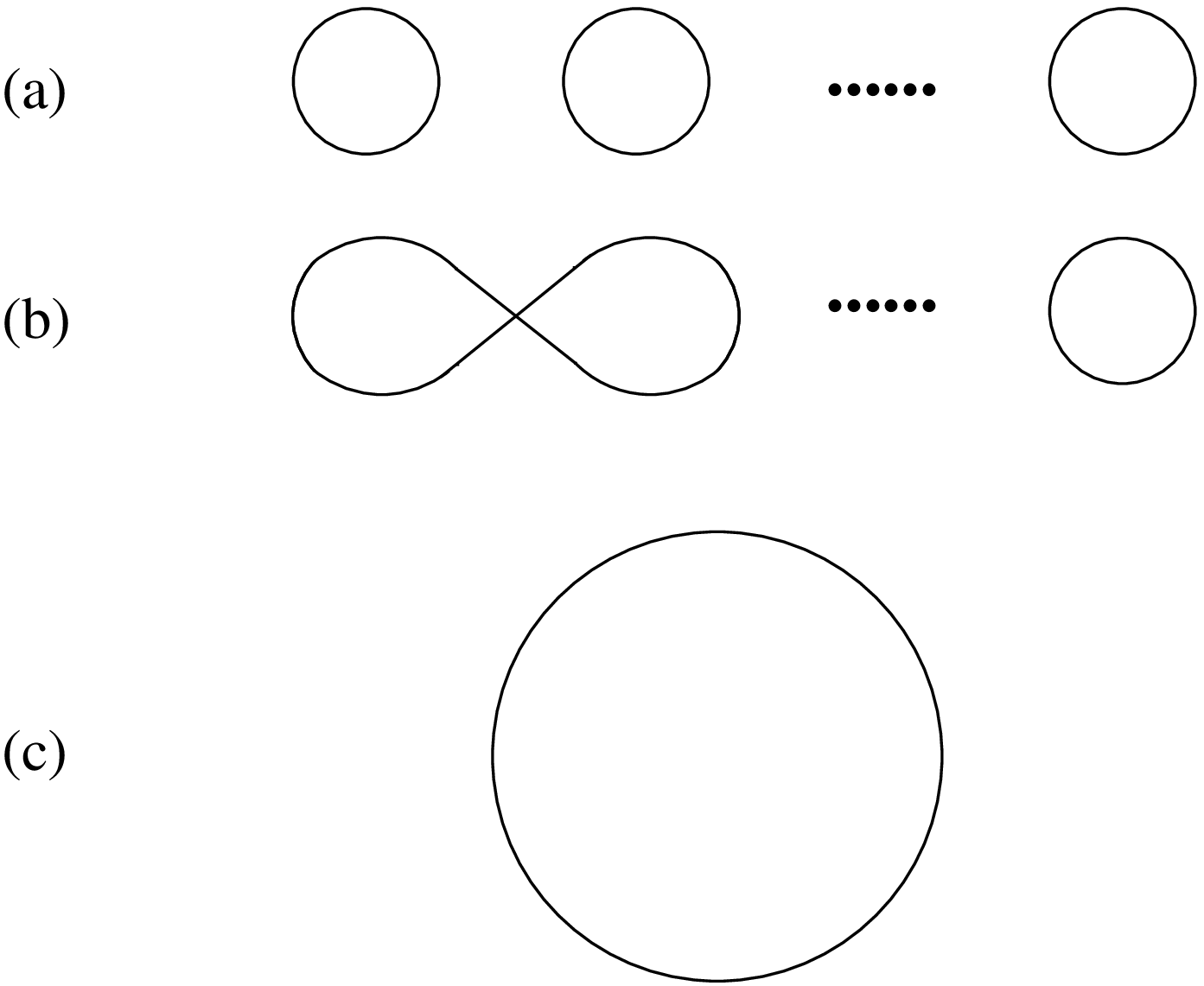}} 

{\bf Figure 1.}{\ \it
Boundary conditions for a family of disconected 
intervals in $\R$ might correspond to different topologies. In 
case (a) we have N disconnected circles. In case (b) two circles merge
into an {\rm eight}. The last case (c) corresponds to a single circle.
Generic boundary conditions describe $N$ open disconnected strings}
\medskip

If we have several disconnected intervals $M=\cup _{i=1}^N[a_i,b_i]$
then the space of boundary conditions is given by $U(2N)$. If we identify
some of the boundaries with opposite orientations we can generate different
 closed manifolds with $r\leq N$ circle components. The corresponding
 operators go from
\eqn\sper{U_1=\pmatrix{0&\I_N\cr
\I_N&0}}
for the connected manifold, till
\eqn\ssper{U_N=\pmatrix{0&1&0&0&0 &\cdots  &0 & 0\cr
1&0&0&0&0 &\cdots  & 0& 0\cr
0&0&0&1&0 &\cdots  & 0& 0\cr
0&0&1&0&0 &\cdots  & 0& 0\cr
\cdot &\cdot&\cdot&\cdot&\cdot &\cdots  & \cdot& \cdot\cr
0&0&0&0&0 &\cdots  & 0& 1\cr
0&0&0&0&0 &\cdots  & 1& 0\cr}}
the  manifold made of $N$ disconnected circles (see Fig. 1).
Therefore, in $U(2N)$ it is possible to describe an smooth transition
from one topology to the other, which provides a theoretical framework
for topological transitions \rbal, in particular, for a 
joint description of open and closed strings.

\appendix{B}{Higher dimensional boundary conditions.}

The last type of boundary condition can be generalized for arbitrary
dimensions.
Let us first consider the 2-dimensional disk $D=\{x\in \R^2; ||x||\leq 1 \}$.
The space  of all boundary conditions of the laplacian can then be identified
with $\CU(L^2(S^1))$  which in this case is an infinite dimensional 
group.
Examples of selfadjoint extensions include  Dirichlet
$U=-\I$ and   Neumann $U=\I$ boundary conditions. There are other boundary
conditions  which correspond to topological foldings of the disk into  
Riemann surfaces  any genus.
Let us first consider the case of the sphere $S^2$.
\eqn\sph{\eqalign{&\partial_\theta \varphi=0\cr
 \int_{S^1} &\dot\varphi \, d\mu_\Gamma=0}}
The associated unitary operator is
\eqn\usph{U_{0}=\pmatrix{ P_0 &  0\cr
 0&  \I-P_0 } }
where $$P_0={1\over 2\pi}\int_{S^1} \cdot\, d\theta $$ is the projector to
the constant functions subspace $\CH_0=\{\varphi\in L^2(S^1);
\partial_\theta \varphi =0\}$ of $L^2(S^1)$ and
we have split $L^2(S^1)$ into $\CH_0$ and its orthogonal complement
$\CH_0^\dagger$. The selfadjoint extension $U_{0}$
corresponds to the topology of the sphere $S^2$.

      It is obvious how to generalize the above construction for higher genus.
For genus 1 surfaces we decompose  the boundary $S^1$ into its four quadrants
$I_1=[0,\pi/2]$, $I_2=(\pi/2,\pi)$, $I_3=[\pi,3\pi/2]$ and $I_4=(3\pi/2,2\pi)$
and identify the points $\theta\in I_1$ with   $3\pi/2-\theta\in I_3$,
and the points $\theta'\in I_2$ with  $5\pi/2-\theta'\in I_4$.
If we split $L^2(S^1)=\oplus_{r=1}^4 L^2(I_r)$ into
its components over the four quarters of $S^1$. The corresponding selfadjoint
extension is given by the $U(8)$unitary operator
\eqn\utor{U_{1}=\pmatrix{ 0   &0& \I_2 &0\cr
 0 & 0 & 0 & \I_2 \cr
\I_2 & 0 & 0 & 0  \cr
 0 &\I_2 & 0 & 0} }
defines the quantum selfadjoint extension of $\H$ which corresponds to toroidal
compactification of $M$.

Other splittings of the circle give rise to different tori, but
for all of them it is necessary to have an isometry between the pairs of
alternating arcs.

For arbitrary genus $g$, we have the straightforward generalization, via
an splitting of the circle $S^1$ into $2g+2$ arcs. The unitary operator
\eqn\utor{U_{g}=\pmatrix{0   &\I_{2g+2}\cr
 \I_{2g+2}& 0} .}
In this way all possible string world sheets transitions
can be described in the set of boundary conditions $\CU (L^2(S^1))$ in a smooth
set up. If we substitute the identity operators $\I_N$ by diagonal
phases the boundary condition describes the effect of magnetic
fluxes crossing though the handles of the compact surface of genus $g$.
Moreover the creation and annihilation of bubbles, baby universes
and transition between them can also be described
by considering families of disks in analogy with the one-dimensional case
of Appendix A.

 Many of those boundary conditions involving non-trivial topological
foldings can be generalized for the higher dimensional case. However much
more  conditions associated to different topological manifolds appear.
The simplest one is the corresponding to the folding of the $n$-dimensional
ball to the $n$-dimensional sphere $S^n$. The corresponding unitary operator
is given by
\eqn\nsph{U_{S^n}=\pmatrix{ P_0 &  0\cr
 0&  P_0- \I } }
where as in the one-dimensional case
$$P_0={1\over 2\pi}\int_{S^n} \cdot \ d\mu_{S^n} $$ denotes the projector to
the  subspace of constant functions $\CH_0=\{\varphi\in L^2(S^n);
d \varphi =0\}$  and
 $L^2(S^n)=\CH_0\oplus \CH_0^\dagger$ has been split  into $\CH_0$
and its orthogonal complement $\CH_0^\dagger$.

The common feature of all these boundary conditions involving topology change
is that the unitary matrix have pairs of eigenvalues $ (1,-1)$ indicating
that all of them belong to the Cayley manifold $\CC_-\cap \CC_+$.

Another type of \bc\ can appear from the choice
$$U={\I\pm i \Delta_\Gamma\over \I\mp i \Delta_\Gamma }.$$ In this case
$\varphi=\pm \Delta_\Gamma
\dot\varphi$, and
$$\left(d\ \Psi_1,d\ \Psi_2\right) =
\left(\Psi_1,\Psi_2\right)\pm  \left(\varphi_1, \Delta_\Gamma
\varphi_2\right).$$
we might have negative energy levels. In the presence of magnetic field
a similar boundary condition leads to negative energy edge levels
which are also present in the Hall effect in a Corbino disk \sr.

\hskip 3cm \epsfxsize=6cm\epsfbox{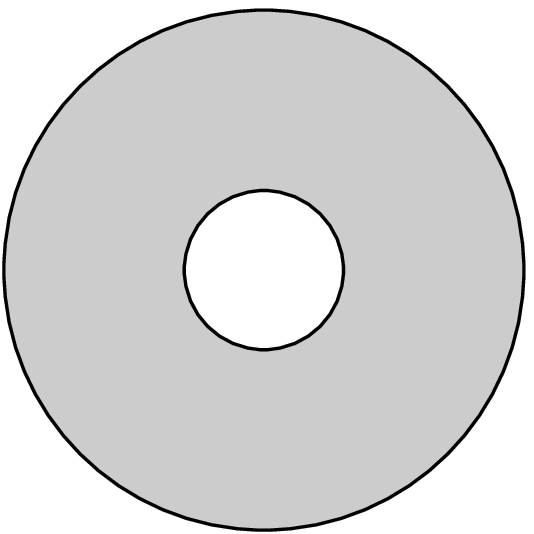}

\centerline{\bf Figure 2. \it The Corbino disk.}

\appendix{C}{ Self-Adjoint Extensions of Higher Order Operators.}

The theory of selfadjoint extensions of Laplace-Beltrami operators developed
throughout the paper can be generalized in a straightforward way for other
differential operators like the  Dirac operator $\Dsl_A$ \ref\rdf{ X. Dai and
D.S. Freed, J. Math. Phys. {\bf 35}(1994) 5155; Erratum {\bf 42}(2001) 2343
} and different powers and
products of $\Dsl_A$ and $\Delta_A^2$. To illustrate how this can
be achieved let us consider for simplicity the  forth order differential
operator given by the square of the covariant Laplace-Beltrami $\Delta_A^2$.
Some of  boundary conditions of $\Delta_A^2$ are induced by those of
$\Delta_A$, but  the space of \bc\ of the square operator is much larger.

The operator $\Delta_A^2$ is symmetric  on
$C^\infty_0 (M,E)$ with respect to the hermitian product
$ \langle .\, , .\, 
\rangle $ of $E$ defined by \riemp,  but in order to obtain
a selfadjoint extension we have to define a larger domain where
 the boundary terms arising from  the integration by parts formula
\eqn\simett{ \langle \psi_1, \Delta_A^2 \psi_2\rangle =
\langle \Delta_A^2\psi_1, \psi_2\rangle + i  \Sigma_2\left(\psi_1,
\psi_2\right),}
vanish. This boundary term
$$\eqalign{ \Sigma_2\left(\psi_1, \psi_2\right) = &
i \int_M\! d \left[
\left( \ast d_{A}\psi_1 ,\Delta_A \psi_2 \right) +
\left( \ast d_{A}\Delta_A \psi_1 , \psi_2 \right) \right.
- \left.
\left( \Delta_A\psi_1 ,\ast d_{A} \psi_2 \right)-
\left( \psi_1 ,\ast d_{A} \Delta_A \psi_2 \right)\right]\cr }$$
only really depends on the boundary values 
$\varphi,\dot\varphi,\Delta_A\varphi,\dot{\Delta_A\varphi}$ defined by
$$\varphi_i=j^\ast \psi_i,\,\dot\varphi_1 \,
d\mu_\Gamma =j^\ast [\ast d_{A}\psi_i];\,
\Delta_{A}\varphi_i=j^\ast [\Delta_{A}\psi_i],\,
\dot{\Delta_{A}\varphi_i}\, d\mu_\Gamma =j^\ast[ \ast d_{A}
\Delta_{A}\psi_i],$$
for i=1,2,
because by Stokes theorem,
$$ \eqalign{\Sigma_2\left(\psi_1, \psi_2\right)=& i
\int_{\Gamma} \,
{j}^\ast\left[\left( \ast d_{A}\psi_1 ,\Delta_{A}\psi_2\right)  +
\left(\ast d_{A}\Delta_A \psi_1 , \psi_2 \right) 
- \left( \Delta_A\psi_1, \ast d_{A} \psi_2\right) -
 \left(\psi_1, \ast d_{A} \Delta_{A}\psi_2\right)  \right]\cr
=&{i} \int_{\Gamma} \,
\left[(\dot\varphi_1 , \Delta_{A}\varphi_2 ) + (\dot{\Delta_{A}\varphi}_1 , 
\varphi_2 )-
(\varphi_1 ,\dot{\Delta_{A}\varphi}_2 )-
(\Delta_{A}\varphi_1 ,\dot\varphi_2 )\right] d\mu_\Gamma.
}$$
Again this boundary term $\Sigma_2(\psi_1,\psi_2)$  measures the net
probability flux across the
boundary. If the operator $\Delta^2_{A}$ were self-adjoint this flux
would have to be balanced or in other words,  $\exp it\Delta^2_{A}$
would be an unitary operator and the  probability preserved.

 The different
self-adjoint extensions of $\Delta_{A}^2$ will be defined in the linear
dense subspaces of $L^2(M, \IC^N)$ containing  $C^\infty_0 (M,E)$
where the boundary term $\Sigma_2$ vanishes. They are therefore 
characterized
by the maximal isotropic subspaces of  the boundary data  space
$$[L^2(\Gamma,\IC^N)]^4\times [L^2(\Gamma,
\IC^N)]^4=\{(\varphi_1,\Delta_{A}\varphi_1,\dot\varphi_1,
\dot{\Delta_{A}\varphi}_1);(\varphi_2,\Delta_{A}\varphi_2,\dot\varphi_2,
\dot{\Delta_{A}\varphi}_2)\}$$  with respect to
the pseudo-hermitian structure $\Sigma_2$.

They are easier
characterized after a double Cayley transform $$C:[L^2(\Gamma,\IC^N)]^4
\times [L^2(\Gamma,
\IC^N)]^4\longrightarrow [L^2(\Gamma,\IC^N)]^4\times [L^2(\Gamma,
\IC^N)]^4$$
 is performed
\eqn\cayy{\eqalign {C&\left((\varphi_1,\dot\varphi_1,\right. 
\left.\Delta_{A}
\varphi_1,\dot{\Delta_{A}\varphi}_1),
(\varphi_2,\dot\varphi_2,\Delta_{A}\varphi_2,
\dot{\Delta_{A}\varphi}_2)\right)\cr
&=((\phi^+_1,\phi^-_1,\Delta_{A}\phi^+_1,\Delta_{A}\phi^-_1),
(\phi^+_2,\phi^-_2,\Delta_{A}\phi^+_2,\Delta_{A}\phi^-_2))\cr
&=\left((\varphi_1+i\dot{\Delta_{A}\varphi}_1,
\varphi_1-i\dot{\Delta_{A}\varphi}_1,
\Delta_{A}\varphi_1+i\dot{\varphi}_1,\Delta_{A}\varphi_1-i
\dot{\varphi}_1)\right. ,\cr
&\left.\phantom{=(}(\varphi_2+i\dot{\Delta_{A}\varphi}_2,
\varphi_2-i\dot{\Delta_{A}\varphi}_2,
\Delta_{A}\varphi_2+i\dot{\varphi}_2,\Delta_{A}\varphi_2-i
\dot{\varphi}_2)
\right), \cr}}
 because
the maximal isotropic spaces with respect to the  transformed product
\eqn\cayccc{\eqalign{&\Sigma^{c}_2 \left(
(\phi^+_1,\phi^-_1,\Delta_{A}\phi^+_1,\Delta_{A}\phi^-_1),(\phi^+_2,\phi^-_2,\Delta_{A}\phi^+_2,\Delta_{A}\phi^-_2)
\right)\cr
&=\langle\phi^+_1,\phi^+_2\rangle-\langle\phi^-_1,\phi^-_2\rangle
+\langle\Delta_{A}\phi^+_1,\Delta_{A}\phi^+_2\rangle-\langle
\Delta_{A}\phi^-_1,\Delta_{A}\phi^-_2\rangle.}}
are in one--to-- one correspondence with
the space of unitary operators $$U\left(L^2(\Gamma,\IC^N)
\times L^2(\Gamma,\IC^N)\right)$$
of the boundary
space $L^2(\Gamma,\IC^N)\times L^2(\Gamma,\IC^N)$. Therefore, we have proved
the following result.

 \noindent{\bf Proposition:}
{\it The set of self-adjoint extensions of
$\Delta^2_{A}$ is in one-to-one correspondence with the group of unitary
operators of $L^2(\Gamma,\IC^N)\times L^2(\Gamma,\IC^N)$. The subgroup
of unitary operators of the form ${\bf U}=(U,U)$ with $U\in
L^2(\Gamma,\IC^N)$ correspond to the \bc\ induced by those of the second
order operator $\Delta_{A}$}.

\listrefs

 \end